\documentclass[prb,superscriptaddress]{revtex4}

\pdfoutput=1

\usepackage{graphicx}
\usepackage{amsmath, amssymb, amsfonts,bm}
\usepackage{latexsym}
\usepackage{bbm}
\usepackage[protrusion=true, expansion=true]{microtype}
\usepackage[colorlinks]{hyperref}
\begin{document}
\title{St\"uckelberg interference in a superconducting qubit under periodic latching modulation}

\author{M.~P. Silveri}
\affiliation{Department of Physics, Yale University, New Haven, Connecticut 06520, USA}
\affiliation{Department of Physics, University of Oulu, P.O. Box 3000, FI-90014, Finland}
\author{K.~S. Kumar}
\affiliation{Low Temperature Laboratory, Department of Applied Physics, Aalto University School of Science, P.O. Box 15100, FI-00076 AALTO, Finland}
\author{J.~Tuorila}
\affiliation{Department of Physics, University of Oulu, P.O. Box 3000, FI-90014, Finland}
\author{J. Li}\altaffiliation{Present address: Electronic and Nanoscale Engineering, School of Engineering, University of Glasgow, G12 8LT, UK}
\affiliation{Low Temperature Laboratory, Department of Applied Physics, Aalto University School of Science, P.O. Box 15100, FI-00076 AALTO, Finland}
\author{A. Veps\"al\"ainen}
\affiliation{Low Temperature Laboratory, Department of Applied Physics, Aalto University School of Science, P.O. Box 15100, FI-00076 AALTO, Finland}
\author{E.~V. Thuneberg}
\affiliation{Department of Physics, University of Oulu, P.O. Box 3000, FI-90014, Finland}
\author{G.~S. Paraoanu}
\affiliation{Low Temperature Laboratory, Department of Applied Physics, Aalto University School of Science, P.O. Box 15100, FI-00076 AALTO, Finland}

 \date{\today}
\begin{abstract}
When the level separation of a qubit is modulated periodically across an avoided crossing, tunneling to the excited state - and consequently Landau-Zener-St\"uckelberg interference - can occur. The types of modulation studied so far correspond to a continuous change of the level separation. Here we study periodic latching modulation, in which the level separation is switched abruptly between two values and is kept constant otherwise. In this case, the conventional approach based on the asymptotic Landau-Zener (LZ) formula for transition probabilities is not applicable. We develop a novel adiabatic-impulse model for the evolution of the system and derive the resonance conditions. Additionally, we derive analytical results based on the rotating-wave approximation (RWA).
The adiabatic-impulse model and the RWA results are compared with those of a full numerical simulation.
These theoretical predictions are tested in an experimental setup consisting of a transmon whose flux bias is modulated with a square wave form. A rich spectrum is observed, with distinctive features correspoding to two regimes:
slow-modulation and fast-modulation. These experimental results are shown to be in very good agreement with the theoretical models. Also, differences with respect to the well known case of sinusoidal modulation are discussed, both theoretically and experimentally.
\end{abstract}

\maketitle

\section{Introduction}

A paradigmatic example of quantum mechanical time-evolution is the Landau-Zener (LZ) problem~\cite{dau}:  in its modern formulation, a qubit is swept across an avoided crossing of the adiabatic energy states. The model is characterized by the asymptotic LZ probability $p_{\rm LZ}$ of making a transition between the states, which is typically calculated for energy sweeps linear in time. In a coherent system, if these traversals across the crossing are repeated periodically,  one observes the Landau-Zener-St\"uckelberg (LZS) oscillations of the qubit population, caused by interference of the different evolutionary paths~\cite{Shevchenko10}.

LZS interference has been realized in a variety of systems, such as Rydberg atoms~\cite{atoms}, superconducting qubits~\cite{exp}, semiconductor quantum dots~\cite{petta,Ludwig}, donors in silicon nanowires~\cite{franceschi}, nitrogen vacancy (NV) centers in diamond~\cite{du}, nanomechanical oscillators~\cite{weig}, and ultracold atoms in accelerated optical lattices~\cite{opticallattice}. In these experimental realizations, the periodic modulation between two extrema of the transition energy has been achieved by driving the qubit longitudinally with a triangular or sinusoidal signal. By assuming that the extrema are sufficiently far away from the crossing, one can estimate that the transition occurs at the avoided crossing and the transition probability amplitude can be approximated by the asymptotic LZ probability.

However, the LZ method for calculating the transition probability is not applicable when the speed of the sweep is increased. With increasing speed, the transition is no longer located strictly at the avoided crossing, but instead it is spread over a larger energy range. This can be demonstrated in a qubit whose level separation is changed abruptly. Applying naively the asymptotic LZ formula would give $p_{\rm LZ} = 1$, predicting the disappearance of the characteristic interference pattern for repeated traversals. What happens instead is that for a sudden switch the energy range of the LZ transition diverges, meaning that the end points are always within the transition region, and thus the asymptotic LZ formula is not applicable.

In this paper we study a qubit whose energy level separation is switched periodically between two constant values, see Fig.~\ref{scheme} (a).
We call this type of modulation `periodic latching' because in-between the switches the qubit is latched onto a fixed value of the energy separation. In this case we can separate two relevant time scales, fast and slow, for switching and latching, respectively. This brings in a qualitatively new conceptual aspect compared to the sinusoidal or triangular modulation where only one timescale (the period) exists for both the transition and the adiabatic evolution. Moreover, periodic latching results in interference patterns with specific features, qualitatively different from those obtained by sinusoidal or triangular modulation. We will refer to these effects
generically as St\"uckelberg interference, to emphasize the more general character
with respect to the standard LZS interference, where the asymptotic LZ model for transition probabilities is assumed to be valid.

The problem of discontinuous periodic modulation appeared for the first time in nuclear magnetic resonance experiments where the nuclear spin evolution was manipulated by periodic trains of sharp, intense pulses~\cite{waugh}. Recently, the problem of qubit modulation with multiple timescales has attracted renewed attention. For example, Ref.~\onlinecite{kohler} considered modulations with two different Fourier components (at frequencies $\Omega$ and $2\Omega$, or at $\Omega$ and $3\Omega$). Also, aperiodic sequences of sharp pulses have been employed in superconducting circuits to create quantum simulations~\cite{paraoanu} of weak localization of electrons in disordered conductors \cite{cleland} and motional averaging \cite{motional}, while bi-harmonic modulation has been employed to simulate universal quantum fluctuations \cite{simon}. The effect of sudden changes in the energy level separation is similar to that produced by defects and two-level fluctuators~\cite{Makhlin,1overf,ress}. To the best of our knowledge, the periodic latching modulation has not been previously discussed in the literature.

The periodic latching modulation can be implemented in a circuit-QED setup~\cite{transmon}, allowing us to test the theoretical predictions against the experiment. The magnetic flux threading the SQUID of the  transmon~\cite{transmon} can be modulated with a square pulse pattern, which naturally brings in two time scales: the duty cycle provides the periodicity, while the raise and fall times occur on a different, much shorter, time-scale.
However, the transition frequencies in this setup lie in the GHz range. This presents a technical challenge for the realization of the square pulses, which even with state-of-the art equipment cannot be generated and transmitted undistorted in a cryogenic setup at such high frequencies. We demonstrate that this problem can be circumvented by driving the qubit near resonance which effectively leads, in the rotating frame, to transition frequencies in the MHz range. Rapid and precise control of the qubit's transition frequency is generally important in, {\it e.g.}, the field of quantum computing~\cite{onoff}, and even in the study of quantum fields in curved spacetimes~\cite{Jacobson, DCE, olsonralph, unruhdewitt}. Our results can be seen as step along this line of research, and suggest that the use of a rotating frame could be an
alternative route to
realizing these experiments.

The paper is organized in the following way. In Sec.~\ref{sec:theory}, we construct an adiabatic-impulse theory appropriate for the modeling of periodic latching modulations. We also derive the excited state population in the steady state and the locations of the population extrema. Section~\ref{sec:experiment} is devoted to our experimental realization consisting of a transmon with a flux modulation of square wave form. We show that by dressing the transmon with an additional microwave drive, the resulting effective Hamiltonian is of the generic avoided-crossing form. In Sec.~\ref{results} we compare the experimental data with the numerical results for the transmon, including higher energy levels, and we
discuss the slow-modulation and the fast-modulation regimes. In this section we also develop an analytical model based on the  rotating-wave approximation (RWA). Also, we compare the experimental and numerical data with those resulting from sinusoidal driving, and extract the sideband traces demonstrating the differences in the two forms of driving. Sec.~\ref{sec:conclusions} concludes the paper with a summary and future prospects.

\section{St\"uckelberg interference under periodic latching modulation}
\label{sec:theory}

Conventionally, the periodic level-crossing problem has been studied in terms of the generic
Hamiltonian
\begin{equation}\label{eq:lzs}
\hat{H} (t)=\frac{\hbar}{2}\big[\nu+ f(t)\big]\hat{\sigma}_z+\frac{\hbar g}{2}\hat{\sigma}_x.
\end{equation}
This represents a quantum mechanical two-level system (qubit) with off-diagonal coupling $g$, and diabatic energy level separation $\hbar\nu$ modulated by a time-periodic function $f(t)$. The types of modulation $f(t)$ studied so far have been of sinusoidal and triangular form~\cite{atoms, exp, petta, Ludwig, franceschi, du, weig, opticallattice}. The latter case corresponds precisely to the linear time-dependence originally introduced by Landau~\cite{dau}, while the former can be approximated as linear near the avoided crossing region. In both of the above cases there is only one time scale involved in $f(t)$, that is, the period of the modulation. Accordingly, the dynamics of the time-periodic system is formally discretized into an adiabatic evolution interrupted by instantaneous non-adiabatic Landau-Zener transitions in the close vicinity of the avoided crossing~\cite{Shevchenko10}. The probability of a single transition between the adiabatic energy states is given by the celebrated Landau-Zener formula. Moreover, the periodic Landau-Zener transitions can interfere, leading to LZS-oscillations of the qubit population~\cite{Shevchenko10}.

\subsection{The Landau-Zener approach}

In the case of latching modulation, the transition frequency is instantaneously and periodically switched between two constant values. In contrast to the previous studies, the adiabatic evolution and the transitions are now clearly separated in the time domain. This kind of time-evolution can be achieved by using a modulation $f(t)$ with two very different timescales: one very slow, realizing the
simplest adiabatic evolution in a time-independent form for a time $2\pi/\Omega$ where $\Omega$ is the angular frequency of the modulation; and the other one very fast, corresponding ideally to a sudden change in the frequency of the qubit. Since in-between the sudden transitions the system is `latched' to one of the transition frequencies, we will refer to such modulation as periodic latching modulation. In practice, the latching modulation can be created with a square wave function (50\% duty cycle) with amplitude $\delta$,
\begin{equation}
f_{\rm sq}(t) = \delta\ {\rm sgn}[\cos(\Omega t)]. \label{square_pulses}
\end{equation}
Let us recall the Landau-Zener formula:
\begin{equation}
p_{\rm LZ} = \exp \left[-\frac{\pi}{2}\frac{g^2}{|v_{\rm LZ}|}\right],
\end{equation}
where the rate of change of the diabatic energy separation evaluated at the crossing is given by $v_{\rm LZ} = [df (t)/dt]_{\rm cross}$. A direct application of this formula for the case of ideal sudden latching would correspond to an infinite LZ speed $v_{\rm LZ} = \infty$. Then the Landau-Zener formula yields $p_{\rm LZ}=1$, predicting the absence of interference and therefore constant qubit population~\cite{Shevchenko10}. The reason for the inadequacy of the LZ approach for the problem of latching is that this formalism requires the asymptotic match of the adiabatic and diabatic states far enough from the avoided crossing point. In our case this assumption is not satisfied, since $\delta$ can be such that the latching points are inside the transition region.
The problem of the width of the transition and the validity of the LZ approximation has been studied previously in the literature. It has been shown \cite{Shevchenko10,Mullen}
that if the LZ speed increases to values $v_{\rm LZ} > g^2/4$ then the transition time scales as $t_{\rm LZ} \sim 1/\sqrt{v_{\rm LZ}}$. As a result, the width of the transition region increases as $v_{\rm LZ}
t_{\rm LZ} \sim \sqrt{v_{\rm LZ}}$. In our case it can easily exceed the range of the extreme points of the latching modulation, when $v_{\rm LZ}\gtrsim 4(\nu\pm\delta )^2$.

Conversely,
if one wishes to study the parameter regime outside the region of validity of the LZ formula, the sinusoidal modulation is not the optimal choice.
To reach a large LZ speed with a modulation $f_{\rm sin}(t) = \delta\sin (\Omega t)$ we would need to increase $\Omega$ to values above the modulation amplitude $\delta$, as can be checked from the criterion above. However, for $\Omega \gg \delta$ motional averaging washes out most of the features in the spectrum \cite{motional}. This effect of motional averaging is indeed observed in our experiments with sinusoidal modulation described below. In contrast, in the case of latching modulation we have the advantage of using an additional very fast time-scale (the linear ramp time). The fast ramp, together with the finite $\delta$, automatically ensures that we are outside the regime of validity of the LZ formula and therefore frees the slow-time variables $\delta$
and $\Omega$, and therefore frees the slow-time variables d and O for realizing interference.

Thus we have to develop a different approach to calculating the non-asymptotic transition probabilities suitable for the case of periodic latching modulation. This is done next by employing the so-called adiabatic-impulse method.

\subsection{Unitary evolution of a qubit  under periodic latching modulation}

\begin{figure}
\includegraphics[width=0.4\linewidth]{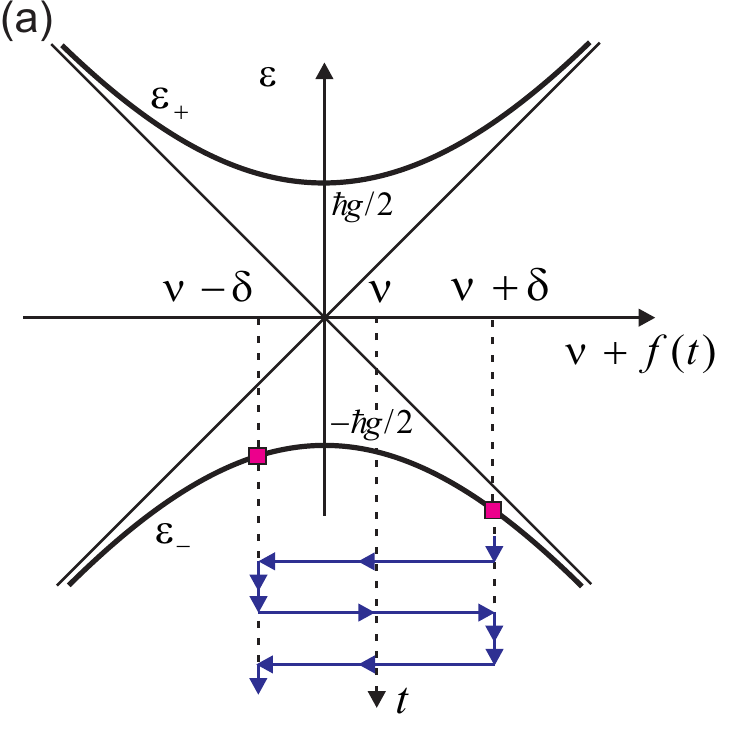}\includegraphics[width=0.42\linewidth]{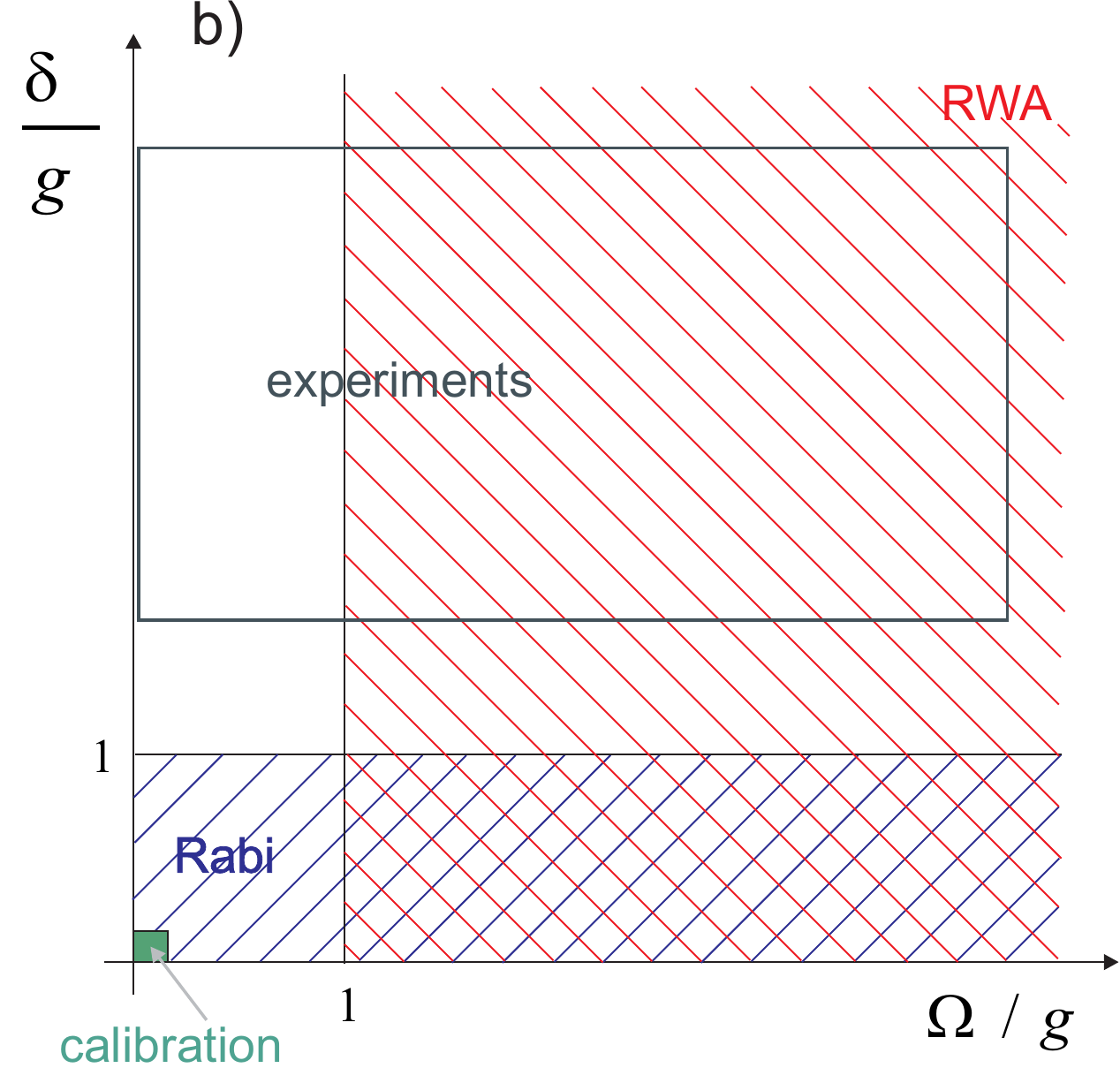}
\caption{(a) Schematic of the avoided crossing, showing qubit energy levels as a function of the instantaneous transition frequencies. The eigenenergies $\epsilon_{\pm}$ are represented with thick solid lines. The blue arrowed lines and the dashed lines illustrate the periodic latching modulation. (b) Diagram of various regimes at different modulation parameters. The range of
 experimentally achievable parameters is shown as a rectangle (drawn smaller for clarity).} \label{scheme}
\end{figure}

Let us consider that the latching period starts from right $(r)$ latch position, where the transition frequency is
$\nu + \delta$. The other latch position is referred to as `left' $(\ell)$ and the corresponding energy gap is $\nu -\delta$, see Fig.~\ref{scheme} (a). The Hamiltonian is diagonalized straightforwardly in both latches, leading to the eigenenergies
\begin{eqnarray}
\epsilon_{\pm}^{(r)} &=& \pm \frac{\hbar}{2} \sqrt{(\nu + \delta)^2 + g^2},\\
\epsilon_{\pm}^{(\ell)} &=& \pm \frac{\hbar}{2} \sqrt{(\nu - \delta)^2 + g^2},
\end{eqnarray}
and the corresponding eigenstates $|\psi_{\pm}^{(r,\ell)}\rangle$, see the Appendix.

We can switch from the eigenbasis of the 'right latch' to that of the 'left' by making a unitary transformation
\begin{equation}
U_{r \rightarrow \ell} \equiv \left(\begin{array}{cc}
\sqrt{1-p_{\rm s}} & \sqrt{p_{\rm s}}\\
-\sqrt{p_{\rm s}} & \sqrt{1-p_{\rm s}}
\end{array}\right), \label{matrixU}
\end{equation}
where $\sqrt{p_{\rm s}}=\langle \psi_+^{(\ell)}|\psi_-^{(r)}\rangle$ and $\sqrt{1-p_{\rm s}}=\langle \psi_-^{(\ell)}|\psi_-^{(r)}\rangle$.  Since $\hat{H}$ is symmetric, $p_{\rm s}$ can be taken to be real, representing the sudden-switch transition probability. Naturally, $U_{\ell \rightarrow r} = U^{-1}_{r \rightarrow \ell} = U^{T}_{r \rightarrow \ell}$. When the system is switched from one latch to the other, we assume that it does not have time to react by adjusting its state. Accordingly, the (instantaneous) unitary time-evolution during the switching is given by $U_{r\rightarrow \ell}$ or $U_{\ell \rightarrow r}$. The validity of this sudden approximation is studied in detail in the Appendix.

In-between the switches, the system is ``parked'' in either of the latches $\ell$ or $r$, and it gathers adiabatic phase in the corresponding eigenbasis:
\begin{equation}
U_{\phi}^{(r, \ell)} \equiv \left(\begin{array}{cc}
e^{-i \phi^{(r,\ell)}}& 0\\
0 & e^{i\phi^{(r,\ell)}}
\end{array}\right),
\end{equation}
where $\phi^{(r, \ell)} \equiv \left[\epsilon_+^{(r, \ell)} - \epsilon_-^{(r, \ell)}\right]\pi /2\hbar \Omega$. 
During one period, the time-evolution of a state $|\Psi(0)\rangle$ starting from the right latch can be written as
\begin{equation}
\left| \Psi\left(\frac{2\pi}{\Omega } \right)\right> = U\bigg(\frac{2\pi}{\Omega}\bigg)|\Psi(0)\rangle,
\end{equation}
where
\begin{equation}
U\bigg(\frac{2\pi}{\Omega}\bigg)\equiv U_{\phi/2}^{(r)}U^T_{r \rightarrow \ell} U_{\phi}^{(\ell )} U_{r \rightarrow \ell} U_{\phi/2}^{(r)} = \left(\begin{array}{cc}
\alpha & -\gamma^* \\
\gamma & \alpha^*
\end{array}\right),
\end{equation}
 and
\begin{eqnarray}
\alpha &=& (1-p_{\rm s}) e^{-i\left[\phi^{(r)}+\phi^{(\ell)}\right]}+p_{\rm s} e^{-i\left[\phi^{(r)}-\phi^{(\ell)}\right]},\\
\gamma &=& -2i\sqrt{p_{\rm s}(1-p_{\rm s})}\sin \phi^{(\ell )}.
\end{eqnarray}
Starting from the ground state $|\Psi(0)\rangle = |\psi^{(r)}_{-}\rangle$, the probability of finding the system in the excited state after one period is given by
\begin{equation}
P_{+} = |\gamma|^2 = 4 p_{\rm s}(1-p_{\rm s})\sin^2 \phi^{(\ell )}. \label{phi1}
\end{equation}
The structure of this equation resembles St\"uckelberg's single-period population in the LZS-model~\cite{Shevchenko10}, with $p_{\rm s}$ playing the role of the Landau-Zener probability. However, unlike the case of conventional continuous modulation, the probability $p_{\rm s}$ does not depend on the frequency of modulation. Moreover, as already pointed out, the asymptotic Landau-Zener probability formula
for linear switching yields the incorrect result $P_{+} =0$, while in our case the probability $p_{\rm s}$ is not necessarily $1$, allowing distinct evolutionary paths that can interfere. The Landau-Zener result can be recovered only in the limit of large driving amplitude $\delta \gg |\omega_0 - \omega|, g$, in which case we can neglect the effects due to $g$, resulting in $p_{\rm s} =1$.


After $n$ periods, one has~\cite{Bychkov}
\begin{equation}
\left[U\bigg(\frac{2\pi}{\Omega}\bigg)\right]^n \equiv \left(\begin{array}{cc}
u_{11}(n) & -u_{21}^*(n) \\
u_{21}(n) & u_{11}^*(n)
\end{array}\right),
\end{equation}
with
\begin{eqnarray}
u_{11}(n) &=& \cos n\phi + i(\textrm{Im}~\alpha)\frac{\sin n\phi}{\sin \phi},\\
u_{21} (n)&=& \gamma \frac{\sin n\phi}{\sin \phi},\\
\cos \phi &=& \textrm{Re}~\alpha.
\end{eqnarray}
Thus, the excited state population after $n$ periods is
\begin{equation}
P_+(n)=|u_{21}(n)|^2 = |\gamma|^2 \frac{\sin^2 n\phi}{\sin^2\phi}.
\end{equation}
By averaging over $n\gg 1$ periods, we obtain the time-averaged excited state population
\begin{equation}
\bar{P}_+ = \frac12 \frac{|\gamma|^2}{1-(\textrm{Re}~\alpha)^2}= \frac12 \frac{|\gamma|^2}{|\gamma|^2 + (\textrm{Im}~\alpha)^2}, \label{p+}
\end{equation}
where $|\gamma|^2+|\alpha|^2 =1$.

\subsection{Resonances}

The maximum excited state population $\bar{P}_+$ , {\it i.e.} a resonance, is obtained when $\textrm{Im} \alpha=0$:
\begin{equation}
(1-p_{\rm s})\sin\left[\phi^{(r)}+\phi^{(\ell)}\right]+p_{\rm s}\sin\left[\phi^{(r)}-\phi^{(\ell)}\right] = 0. \label{resonance}
\end{equation}
This can be analyzed further in the regimes $p_{\rm s} \approx 1$ or $p_{\rm s} \approx 0$, resulting in the conditions
\begin{eqnarray}
\phi^{(\ell)}-\phi^{(r)}&=&m_-\pi , \label{res1}\\
\phi^{(\ell)}+\phi^{(r)}&=&m_+\pi, \label{res2}
\end{eqnarray}
respectively. The first resonance condition is thus valid for relatively large values of $\delta$ compared to $g$ and $|\nu|$, {\it i.e.} $\delta \gtrsim |\nu|$
and $\delta \gg g$ ,
while the second one is valid when $\delta \lesssim |\nu |$ and $|\nu |\gg g$.
It is, however, instructive to plot them in the entire range of $\delta$ (see Fig.~\ref{Square_fig}). Let us note that in the case of sinusoidal or triangular modulation one obtains a resonance condition similar to Eq. (\ref{resonance}) (see Ref.~\onlinecite{ashhabnori}) but, in contrast to those, in our case the validity regimes of Eqs. (\ref{res1}-\ref{res2}) do not depend on the value of the modulation frequency $\Omega$.

In addition to these resonances that refer to the cyclic evolution, we can identify certain resonance and anti-resonance conditions originating from the single-period
population $P_+$ from Eq.~\eqref{phi1}. The single-period resonance is obtained from the maximum of $P_+$,
\begin{equation}
\phi^{(\ell)}  =\frac{2n+1}{2}\pi  . \label{res3}
\end{equation}
If the evolution after one period results in the same initial state, that is $P_+ = 0$, we have anti-resonance or coherent destruction of tunneling \cite{CDT}. The anti-resonance condition can be written as
\begin{equation}
\phi^{(\ell)}= n'\pi.\label{antiresonance}
\end{equation}
In the expressions above, $m_+$ ,$m_-$,$n$, and $n'$ are integers. We note that the single-period resonance and anti-resonance analytic conditions obtained above are shifted from those given by the standard LZS-model for sinusoidal modulation, in agreement with our numerical simulations and with the experimental data presented later. This is because in the case of continuous modulation the system collects a Stokes phase~\cite{Shevchenko10} during the non-adiabatic transitions, whereas in the case of latching modulation it does not. Also, the average steady state population (\ref{p+}) depends on the starting latch. The steady state occupation can be obtained by averaging over all possible initial phases of the modulation pulse~\cite{Tuorila13}. Nevertheless, we are mainly interested on the locations of the resonances, which remain invariant under the averaging.

The theoretical description presented here is valid everywhere in the parameter space $(\delta, \Omega )$ of the latching modulation. However, a number of analytically intuitive results can be obtained when
$\Omega \gg g$, a limit called the rotating-wave approximation regime~\cite{ashhabnori}, see Fig.~\ref{scheme} (b). These results are presented in Section \ref{results}. In the following section we present details of the experimental realization of the above scheme.

\section{Experimental realization}
\label{sec:experiment}
\begin{figure}
\includegraphics[width=0.7\linewidth]{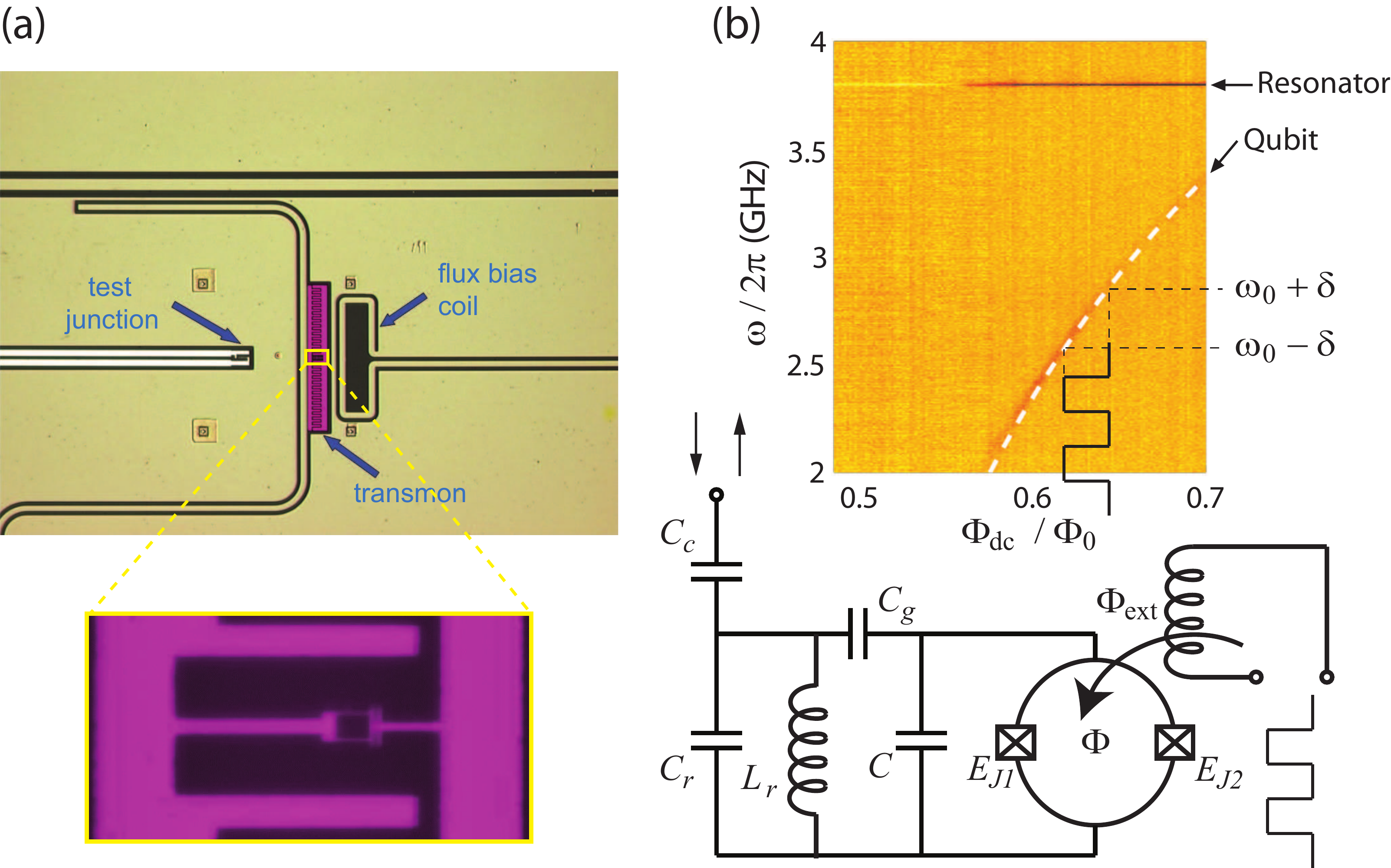}
\caption{\label{fig_transmon} (a) Optical image (artificially colored) of the transmon sample with a detail of the SQUID loop of the transmon. (b) Measured spectrum as a function of the dc-component $\Phi_{\rm dc}$ of the externally-applied magnetic field $\Phi_{\rm ext}(t)$, together with numerical fitting (dashed white line). The lower drawing is a schematic of the equivalent electrical circuit.}
\end{figure}

We have studied the above scheme in the conventional circuit-QED setup~\cite{transmon}, which consists of a capacitively-shunted Cooper pair box (a transmon) coupled to a coplanar waveguide resonator used for dispersive readout. The periodic latching modulation is created by feeding a square pulse current, generated by an arbitrary waveform generator, through the flux bias coil coupled inductively into the SQUID loop used to tune the Josephson energy of the transmon. A schematic of the circuit and an optical image of the sample is presented in Fig.~\ref{fig_transmon}. In the following, we will show that this results in an effective two-level Hamiltonian with periodic latching modulation, thus realizing the Hamiltonian studied in the previous section.

We start with the full Hamiltonian of the transmon, including the coupling with the resonator:
\begin{equation}
\hat{H} = \hbar\omega_r \hat{a}^{\dag}\hat{a} + 2\beta e V_{\rm rms}^0 \hat{n}(\hat{a}^{\dag} + \hat{a})  +4E_{\rm C} (\hat{n} - n_g)^2 - E_{\rm J1}\cos\hat{\varphi}_1 - E_{\rm J2}\cos\hat{\varphi}_2 , \label{eq3_transmon_hamiltonian1}
\end{equation}
where $E_{\rm C} = e^{2}/2C_{\Sigma}$ is the single-electron charging energy, $C_{\Sigma}$ is the total parallel capacitance (including the shunt), $E_{\rm J1}$ and $E_{\rm J2}$ are the Josephson energies of the two Josephson junctions, and $n_{g}$ is the effective offset of the number of Cooper pairs. The resonator frequency is $\omega_{r}=1/\sqrt{L_{r} C_{r}}$ and $\hat{a}$ denotes the annihilation operator of the resonator mode. Also, $V_{\rm rms}^0 = \sqrt{\hbar \omega_{r}/2C_{r}}$ and $\beta=C_{g}/C_{\Sigma}$~\cite{transmon}. In the following, we will concentrate on the bare qubit part consisting of the three last terms in the above Hamiltonian.

The Hamiltonian~(\ref{eq3_transmon_hamiltonian1}) results from the circuit quantization of the qubit coupled to the resonator. As usual,
$\varphi_1$ and $\varphi_2$ denote the gauge-invariant phase differences across the two junctions, and they fulfill the fluxoid quantization condition
\begin{equation}
\varphi_1 - \varphi_2 = 2 \pi \Phi / \Phi_0 \ ({\rm mod} \ 2\pi).
\end{equation}
Here $\Phi_0 = h/2e$ is the flux quantum,
and $\Phi$ is the total magnetic flux through the loop, which is the sum of the external bias flux $\Phi_{\rm ext}$ and the screening flux $\Phi_{\rm s}$. Normally the loop inductance of a transmon is negligibly small, therefore $\Phi \approx \Phi_{\rm ext}$, and, to simplify the notations, we take the flux $\Phi_{\rm ext} \in [-\Phi_{0}/2,\Phi_{0}/2]$. We define $\hat{\varphi} \equiv (\hat{\varphi}_1 + \hat{\varphi}_2)/2$ , $E_{\rm J\Sigma} \equiv E_{\rm J1}+E_{\rm J2}$ and assume that the transmon asymmetry $d \equiv \left(E_{\rm J2} - E_{\rm J1}\right) / E_{\rm J\Sigma} \ll 1$. The transmon is flux-biased at a constant value $\Phi_{\rm dc}$, on top of which we overlap the time-dependent square pulse flux: $\Phi_{\rm ext}(t)=\Phi_{\rm dc} + \Phi_{\rm sq} (t)$. As a result, the transmon part of Hamiltonian~(\ref{eq3_transmon_hamiltonian1}) can be written as
\begin{equation}
\hat{H}_0 = 4E_{\rm C}(\hat{n} - n_g)^2 - E_{\rm J\Sigma} \cos \left(\frac{\pi \Phi_{\rm dc}}{\Phi_0}\right) \cos \hat{\varphi}  + E_{\rm J\Sigma} \sin\left(\frac{\pi\Phi_{\rm dc}}{\Phi_0}\right) \sin\left[\frac{\pi \Phi_{\rm sq}(t)}{\Phi_0}\right] \cos\hat{\varphi}.
\end{equation}

It is convenient to introduce the standard harmonic oscillator creation and annihilation operators $\hat{b}$, $\hat{b}^{\dag}$ associated with the operators $\hat{\varphi}$ and $\hat{n}$,
\begin{eqnarray}
\hat\varphi &=& \varphi_{\rm zpf} \left( \hat b^\dag + \hat b \right), \label{var}\\
\hat n &=& \frac{i}{2\varphi_{\rm zpf}}\left( \hat b^\dag - \hat b \right),\label{en}
\end{eqnarray}
where $\varphi_{\rm zpf}\equiv \left[2E_{\rm C}/[E_{\rm J\Sigma}  \cos(\pi \Phi_{\rm dc} / \Phi_0 )]\right]^{1/4}$.
Accordingly, we have $[\hat{b},\hat{b}^{\dag}]=1$ since $[\hat\varphi, \hat n ] = i$.
In order to minimize the effects of charge fluctuations, the constant flux bias is chosen so that the zero-point phase fluctuations are small,
$\varphi_{\rm zpf} \ll 1$. In this case, the effective offset charge $n_g$ can be eliminated by making a gauge transformation, similar to Ref.~\onlinecite{transmon}. Since the phase is localized with only small fluctuations around the equilibrium position, the local minima of the cosine potential $\cos\varphi$ can be well approximated by a fourth order polynomial.

The Hamiltonian operator of the qubit part is then written as
\begin{widetext}
\begin{equation}
\hat H_{0}  \approx \hbar\omega_p \hat b^\dag \hat b  - \frac{E_{\rm C}}{12}\left( \hat b^\dag + \hat b \right)^4 -\frac{E_{J\Sigma}}{2}\varphi_{\rm zpf}^2\sin\left(\frac{\pi\Phi_{\rm dc}}{\Phi_0}\right)
(\hat b^\dag+\hat b)^2\left[1-\frac{1}{12}\varphi_{\rm zpf}^{2}(\hat b^\dag + \hat b)^2\right]\sin\left[\frac{\pi\Phi_{\rm sq} (t)}{\Phi_0}\right],
\label{eq3_transmon_hamiltonian3}
\end{equation}
\end{widetext}
where the plasma frequency $\omega_p$ is defined as
\begin{equation}
\hbar\omega_p = \sqrt{8E_{\rm C} E_{\rm J\Sigma} \cos\left(\frac{\pi \Phi_{\rm dc}}{\Phi_0}\right) }.\label{plasma}
\end{equation}
The first two terms in Eq.~(\ref{eq3_transmon_hamiltonian3}) comprise the conventional transmon Hamiltonian: a harmonic oscillator with a quartic perturbation. The latter two terms are due to the time-dependent flux modulation. We will show that in the case of square pulse modulation, these terms will produce the periodic latching modulation of the qubit.

In terms of the unperturbed harmonic oscillator states $\{|j\rangle\}$, one obtains  $\langle j | (\hat b + \hat b^\dag)^2|j\rangle = 2j+1$ and
 $\langle j|(\hat{b}+\hat{b}^{\dag})^{4}|j\rangle = 6j^2 + 6 j + 3$; also the even powers of $\hat \varphi$ do not couple states with different parity. By truncating Hamiltonian~(\ref{eq3_transmon_hamiltonian3}) to the Hilbert space spanned by the two lowest energy levels $\{ |0\rangle, |1\rangle \}$, one gets
\begin{equation}
\hat H_{0} = \frac{\hbar}{2}\left[ \omega_0 + f_{\rm sq}(t) \right]\hat \sigma_z, \label{eq3_transmon_hamiltonian4}
\end{equation}
where the transition energy
\begin{equation}
\hbar\omega_{0} = \hbar\omega_p - E_{\rm C}, \label{omega10}
\end{equation}
and the longitudinal drive
\begin{equation}
\hbar f_{\rm sq}(t) = \frac{E_{J\Sigma}}{2}\sin\left(\frac{\pi\Phi_{\rm dc}}{\Phi_0}\right)(\varphi_{\rm zpf}^4-2\varphi_{\rm zpf}^{2})\sin\left[\frac{\pi\Phi_{\rm sq}(t)}{\Phi_0}\right].
\label{eq3_transmon_transverse}
\end{equation}
By comparing with Eq.~(\ref{square_pulses}), we can identify the latching modulation amplitude $\hbar\delta = E_{J\Sigma}\sin(\pi\Phi_{\rm dc}/\Phi_0)(\varphi_{\rm zpf}^4-2\varphi_{\rm zpf}^{2})\sin(\pi\Phi_{\rm sq}/\Phi_0)$, where $\Phi_{\rm sq}$ is the square wave amplitude of the magnetic flux in the transmon SQUID loop.

Besides the flux modulation, the qubit is also driven via the resonator by another microwave field of frequency $\omega$. By neglecting the quantum fluctuations of the resonator, the second term in Hamiltonian~(\ref{eq3_transmon_hamiltonian1}) can be written as
\begin{equation}
\hat{H}_d = \frac{2\beta e V_{\rm rms}^0}{\varphi_{\rm zpf}}\sqrt{n_r}\cos(\omega t) i (\hat b^\dag - \hat b) = \hbar g \cos(\omega t)\hat{\sigma}_y, \label{transmon_drive}
\end{equation}
where $n_r$ is the number of coherent quanta in the resonator, $\omega$ is the driving frequency, and in the latter equality we have made the two-state truncation and defined $\hbar g\equiv (2\beta e V_{\rm rms}^0/\varphi_{\rm zpf})\sqrt{n_r}$. The vacuum Jaynes-Cummings coupling to the first transition is defined by $\hbar g_{0}\equiv \beta e V_{\rm rms}^0/\varphi_{\rm zpf}$, thus $g = 2 g_{0}\sqrt{n_{r}}$.

Next, we transform into a frame rotating at the driving frequency $\omega$ around the $z$-axis, implemented by the unitary transformation $\exp[-i\omega \hat{\sigma}_{z}t/2]$. With an additional rotation $\hat \sigma_y \rightarrow \hat \sigma_x$, we obtain finally the effective Hamiltonian:
\begin{equation}
  \hat{H}_{\rm eff}(t)=\frac{\hbar}{2}\left[\omega_0 - \omega + f_{\rm sq}(t)\right]\hat{\sigma}_z+\frac{\hbar g}{2} \hat{\sigma}_x ,\label{ham.first}
\end{equation}
which defines a $\sigma_x$-coupled qubit with frequency $\nu = \omega_0 - \omega$ modulated by $f_{\rm sq}(t)$. In other words, in our experiment the generic Hamiltonian~(\ref{eq:lzs}) is realized as an effective Hamiltonian in the subspace of dressed states formed by the qubit and the transverse driving field.

The dependence of the energy level separation on the applied external magnetic flux given by Eqs.~(\ref{plasma}) and~(\ref{omega10}) can be used to extract the transmon parameters. We have diagonalized the full transmon Hamiltonian, and by fitting with the measured spectrum (white dashed line in Fig.~\ref{fig_transmon}) we can extract $E_{\rm C}/h  =  0.35$~GHz, $E_{\rm J\Sigma}/h = 8.4$~GHz, and $d\approx 0.1$. The relaxation rate $\Gamma_{1}/2\pi = 1.2$~MHz and the dephasing rate $\Gamma_{2}/2\pi = 3.1$~MHz were obtained by independent characterization measurements \cite{schuster}. The value $g_{0}/2\pi \approx 80$ MHz for the Jaynes-Cummings coupling between the resonator and the transmon was extracted from vacuum Rabi mode splitting data. To allow a good fidelity in the transmission of the square pulse at relatively high frequencies, the qubit was minimally filtered, with the downside of an increased noise level. The readout of the transmon is based on the ac Stark shift of the resonance $\omega_r$ from its bare frequency $\omega_r/2\pi= 3.795$~GHz, resulting in a change in the microwave reflection coefficient $S_{11}$, which is recorded by a vector network analyzer. For each point in the spectra (corresponding to detuning $\omega - \omega_{0}$ and modulation frequency $\Omega$) we average over 70 such measurements.

\section{Results and discussion}
\label{results}
\begin{figure}
\includegraphics[width=0.7\linewidth]{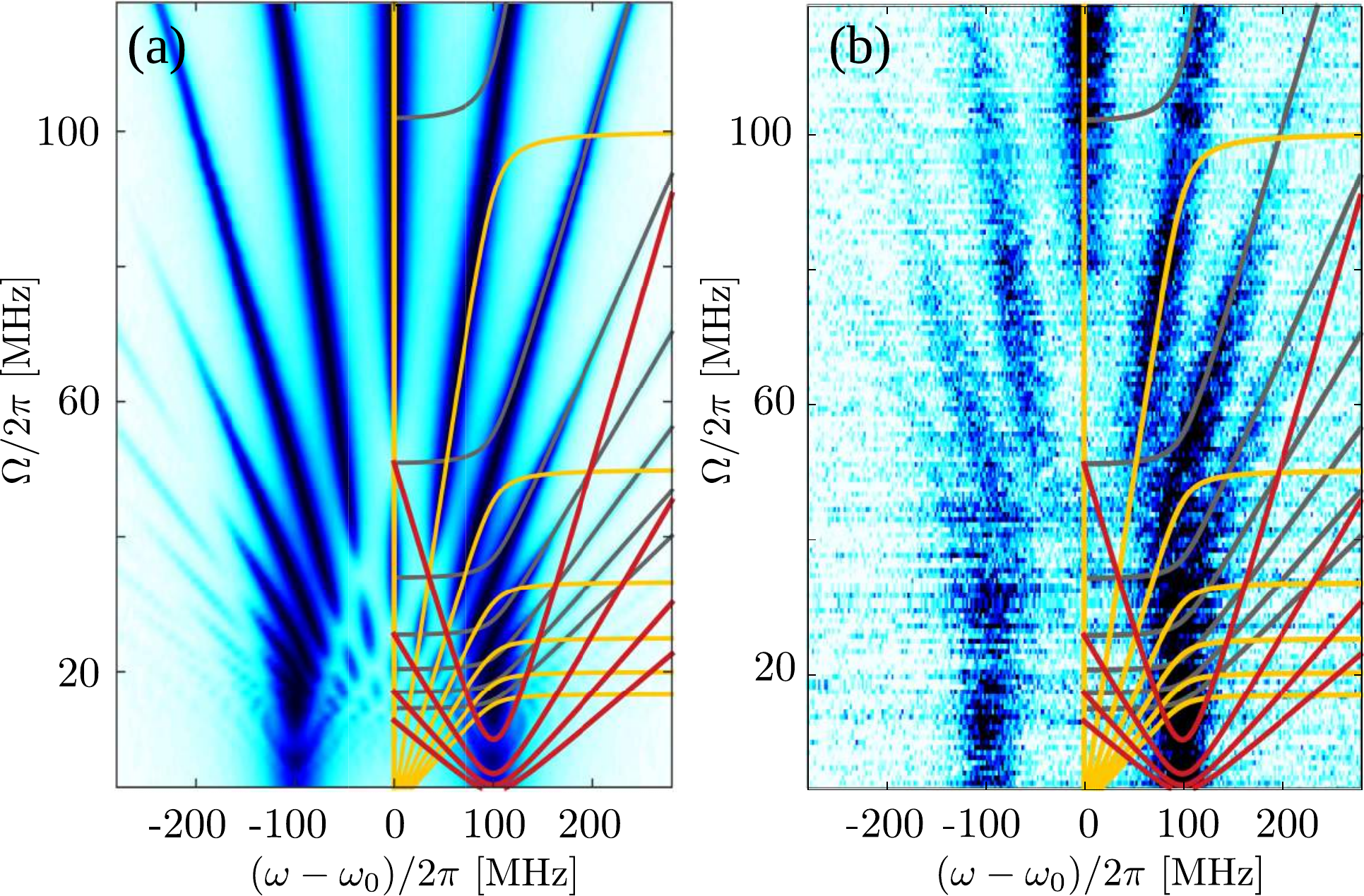}
\caption{\label{Square_fig} St\"uckelberg interference pattern for periodic latching modulation as a function of detuning $\omega-\omega_{0}$ and latching frequency $\Omega$. (a) Simulation of the transmon dispersive shift Eq.~(\ref{shift}) and (b) measured reflection coefficient with parameters $\omega_0/2\pi=2.62$ GHz, $\delta /2\pi= 100$ MHz, $g/2\pi=20$ MHz, $\Gamma_1/2\pi=1.2$ MHz, $\Gamma_2/2\pi=3.1$ MHz, $T=50$ mK. The resonance conditions (\ref{res1}),(\ref{res2}), together with the antiresonance condition (\ref{antiresonance}), are plotted with continuous yellow, grey, and red lines respectively.}
\end{figure}

In this section, we present the experimental results and compare them
with the theoretical predictions, see Fig.\ \ref{Square_fig}. In Fig.\ \ref{Square_fig} (b) we show the experimental
result for the qubit under periodic latching modulation as a function of the
frequency of detuning $\nu /2\pi = (\omega_0-\omega)/ 2\pi$ and the latching frequency $\Omega /2 \pi$. The multi-cycle
analytic resonance conditions Eq. (\ref{res1}) and Eq. (\ref{res2}), in addition to the
the single-period anti-resonance condition of Eq. (\ref{antiresonance}), are overlaid on
top of the right half of the data.
We note that Eq.  (\ref{res1}) (yellow)
and Eq. (\ref{res2}) (gray) explain well the resonances within their validity
ranges: $|\omega-\omega_0|/2\pi \lesssim \delta /2 \pi=100$ MHz,
and $|\omega-\omega_0 | \gtrsim \delta /2 \pi=100$ MHz respectively. The position of
the single-period antiresonance condition Eq. (\ref{antiresonance}) (red) agrees with the
experiment as well, although the comparison in the low-$\Omega$ region is limited by the poor signal-to-noise ratio.
In the following subsections we will provide a more in-depth quantitative comparison between the data and the theory.


\subsection{Comparison with numerical simulations}
\label{comp}

In Fig. \ref{Square_fig} (a) we show the numerical simulations for the latching
modulation together with the resonance conditions resonance conditions (\ref{res1}), (\ref{res2}), and the antiresonance condition (\ref{antiresonance}). The analytic conditions coincide remarkably well, within their range of validity, with the resonances and antiresonances obtained from the numerical results.

\begin{figure}
\includegraphics[width=9.5cm,height=6.5cm]{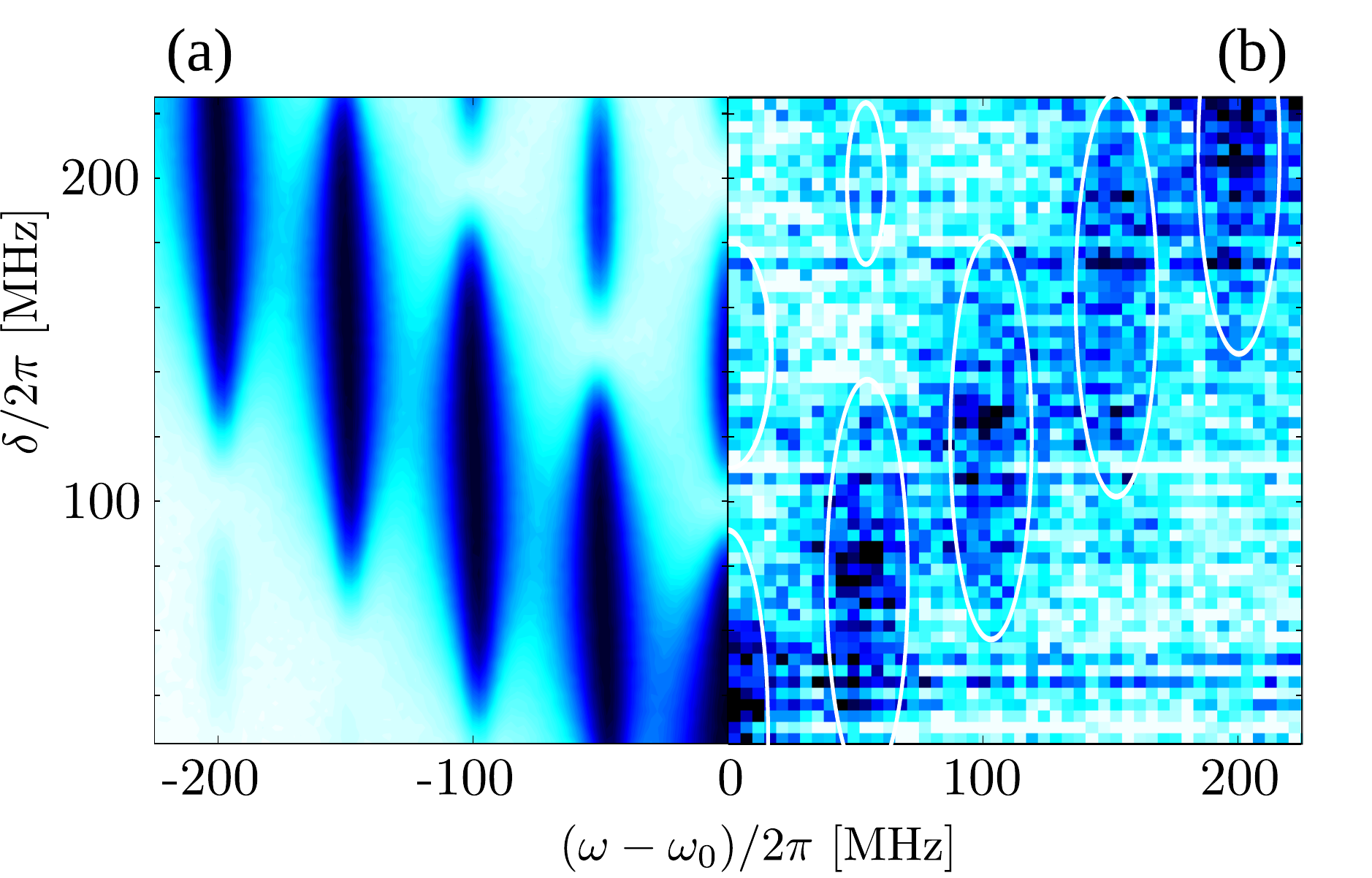}
\caption{\label{LZS} St\"uckelberg interference pattern for periodic latching modulation as a function of detuning $\omega-\omega_{0}$ and latching amplitude $\delta$.
The results are shown in a compact form with the simulation (a) for  negative detunings and experiment (b) for positive detunings. The white ellipses indicate the theoretically-predicted positions of the maxima. The modulation frequency was fixed at $\Omega /2\pi = 50$~MHz and the other parameters are the same as in Fig.~\ref{Square_fig}.
}
\end{figure}

These numerical results are obtained by calculating numerically the steady state population of the driven and modulated transmon in a thermal bath by using the quantum trajectory method~\cite{petruccione}. A transmon is a only weakly anharmonic, therefore in order to take into account the thermal excitations more precisely, the Hamiltonians~\eqref{eq3_transmon_hamiltonian4} and~\eqref{transmon_drive} are extended for the five lowest eigenstates~\cite{transmon,Boissonneault12}. We assume that the driving field couples only the ground state and the first excited state. This approximation is reasonable since the detuning of the field is smaller than the anharmonicity $E_{\rm C}$, that is $\hbar|\omega-\omega_{0}|<E_{\rm C}$. In the dispersive limit~\cite{transmon, Boissonneault12}, the transmon population shifts the eigenfrequency of the resonator by
\begin{equation}
\Delta\omega_{\rm r}=\sum_{i=0}^4 P_{i} \chi_i, \label{shift}
\end{equation}
where $P_i$ is the steady state population of the $i$:th level and $\chi_i=g_{0}^2\left[i/(\omega_{i}-\omega_{i-1}-\omega_{\rm r})-(i+1)/(\omega_{i+1}-\omega_{i}-\omega_{\rm r})\right]$ the state dependent dispersive frequency shift and $\omega_i=i[\omega_0+(1-i)E_{\rm C}/2]$  denote the transmon eigenenergies. In Figs. \ref{Square_fig}-\ref{comparison}, we show the numerical shift $\Delta\omega_{\rm r}$ with respect to the background value calculated for an undriven transmon.

In the simulation we have set the temperature of the environment to $T=50$ mK, which is the base temperature of the refrigerator.  We note however that due to reduced filtering, most likely the noise level felt by the qubit is higher than in the ideal situation. Effective qubit temperatures larger than 100 mK have been determined previously in transmons that do not thermalize properly~\cite{againdicarlo}. Indeed in our simulations we find that by increasing the temperature to higher values results in reduced contrast of the fine structures of Fig.~\ref{Square_fig}(a), in accordance with the experiment. Other sources of non-ideality in the experiment are the presence of mild microwave resonances in the cables and in the sample holder, and the imperfect generation and transmission of square pulses to the qubit.

From Fig. \ref{Square_fig} we can clearly distinguish two regimes, depending on the ratio of the amplitude $\delta$ of the latching modulation and its frequency $\Omega$:
a slow-modulation regime for $\delta /\Omega \gtrsim 2$ and a fast-modulation regime for $\delta/\Omega \lesssim 2$. As we will see, the ratio $\delta /\Omega$ appears as the argument of the sideband amplitudes in the rotating-wave approximation approach developed in Sec. \ref{rotw}. In the slow-modulation regime a fine structure of resonances appear and the differences between the latching modulation and other types of modulation become visible, while in the fast-modulation regime the sidebands are the prominent feature. Note that in the spectrum shown in Fig. \ref{Square_fig} we cover a wide range of values $\delta/\Omega \approx 0.8, ... ,33$ in both the slow-modulation and the fast-modulation frequency regimes.

It is also possible to measure the qubit population as a function of detuning by varying the modulation amplitude at a fixed modulation frequency, which is the standard representation of the LZ interference \cite{Shevchenko10}. To check this, in Fig. \ref{LZS} we plot
a few population oscillations as a function of the detuning $\omega -\omega_{0}$ and latching amplitude $\delta$ at a fixed modulation amplitude $\Omega /2\pi = 50$ MHz. This spectrum covers the parameter range $\delta/\Omega \approx 0.5, ...,4.5$. Although the interference pattern is visible, this representation is not optimal for extracting the differences between sinusoidal and other types of modulation \cite{kohler}.

\subsection{Comparison between periodic latching modulation and sinusoidal modulation}
\label{snn}

\begin{figure}
\includegraphics[width=0.7\linewidth]{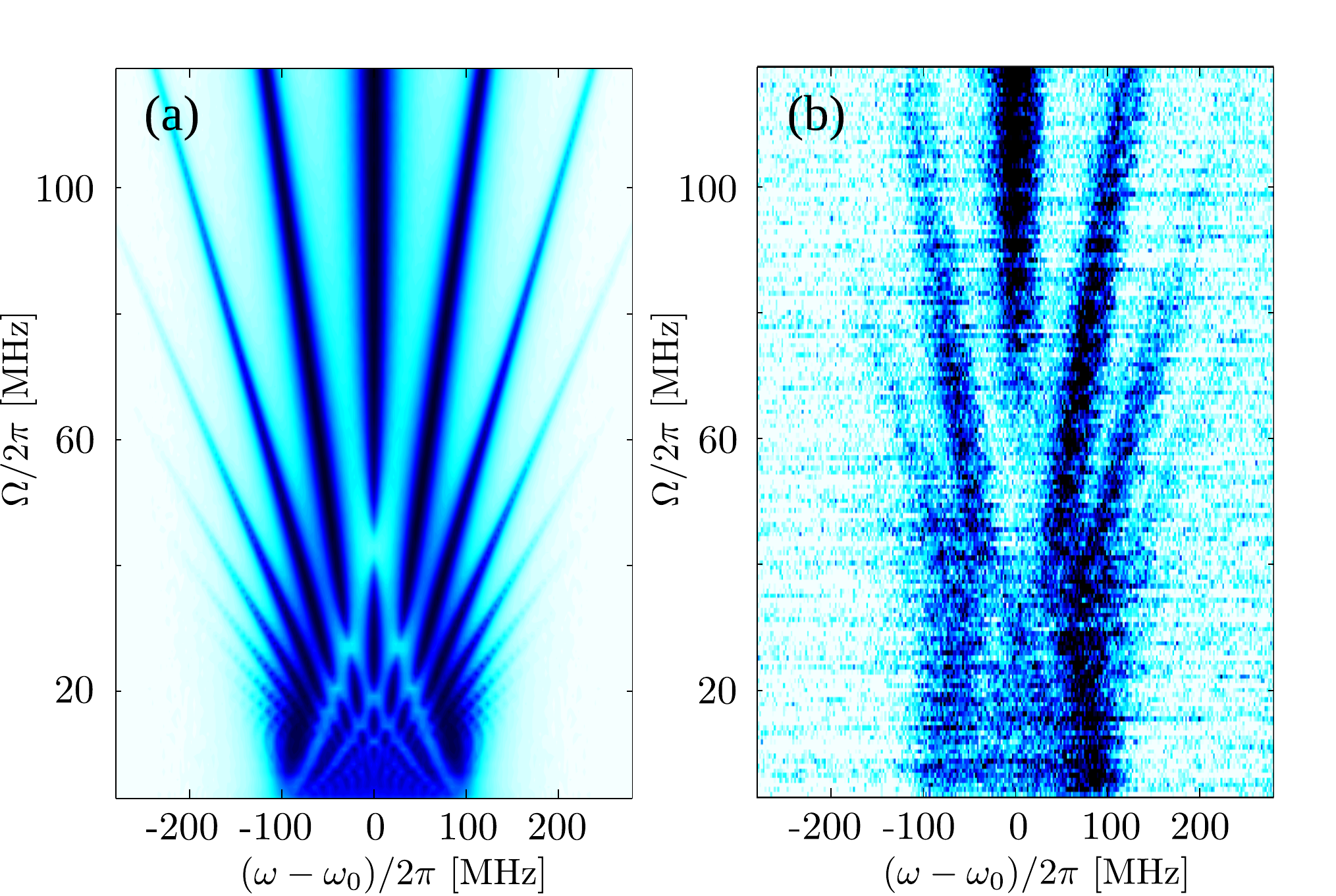}
\caption{\label{Sine_fig}
St\"uckelberg interference pattern for sinusoidal modulation as a function of detuning $\omega-\omega_0$ and modulation frequency $\Omega$. (a) Simulation of the transmon dispersive shift Eq. ~(\ref{shift}) and (b) measured reflection coefficient with sinusoidal modulation of amplitude $\delta /2\pi = 100$ MHz. Other parameters are same as in Fig.~\ref{Square_fig}.}
\end{figure}

We now compare the spectrum of the system under periodic latching modulation with that of a sinusoidal modulation.
 In Fig.\ \ref{Sine_fig} we show the spectrum for sinusoidal modulation with exactly the same parameters (the same qubit frequency $\omega_{0}$ and the same modulation amplitude) as in Fig.\ \ref{Square_fig}. One notices already from the full spectra of Figs.\ \ref{Square_fig} and \ref{Sine_fig} that the structures at low and intermediate modulation frequencies are rather different. To illustrate the differences, in Fig.\ \ref{comparison} (a)
we present a comparison between the sinusoidal and the latching modulation along the second sideband, where for clarity we show the spectra in the low-frequency range, up to $70$ MHz. Here, in order to eliminate the asymmetry seen in the cavity response between positive and negative $\nu = \omega_0 -\omega$, we calculate the average of the sidebands $m=-2$ and $m=2$. The rather poor signal to noise ratio does not allow us to clearly identify all the population oscillations, but some differences can be seen clearly.

Best seen in Fig.~\ref{comparison} (a), the first maximum of the second latching sideband is shifted to higher values of $\Omega$ when compared to that of the sinusoidal modulation.
At very large values of the modulation frequency $\Omega \gg \delta$, both the latching and the sinusoidal-modulation sideband would eventually decrease to zero. In the low-$\Omega$ limit, the dispersive shift due to the sinusoidal modulation remains around $\Delta \omega_r/2\pi \approx 0.9$ MHz when $\Omega$ is decreased, whereas in the case of the latching modulation there is a considerable drop (ideally to  zero). This is due to the fact that for the periodic latching modulation the qubit spends almost no time around $\omega-\omega_{0} = 0$. We have confirmed this behaviour with several other values of $\delta$.

\subsection{Rotating-wave approximation description}
\label{rotw}

In order to get a better understanding of the spectra we develop here an alternative analytic description in terms of the  Hamiltonian~(\ref{ham.first}) and the rotating-wave approximation. This description is valid at relatively large modulation frequencies $\Omega\gtrsim g$, shown schematically in Fig.~\ref{scheme} (b).

We transform the Hamiltonian Eq.~(\ref{ham.first}) into a frame co-rotating with the longitudinal modulation $f(t)$ by employing the unitary transformation
\begin{equation}
\hat{U}(t)=\exp\left(-\frac{\mathrm{i}\hat{\sigma}_z}{2}\int_0^t f(\tau)\mathrm{d}{\tau}\right),
\end{equation}
which, in the Bloch-sphere picture, corresponds to a frame rotating around the $z$-axis with the instantaneous angular velocity $f(t)$. In this frame, the new effective Hamiltonian is obtained from Eq. (\ref{ham.first}) by $\hat{H}'=\hat{U}^\dagger \hat{H}\hat{U}+\mathrm{i}\hbar(\partial_t {\hat{U}}^\dagger) \hat{U}$,
\begin{equation}
\hat{H}_{\rm eff}(t)=\frac{\hbar}{2}(\omega_0-\omega ) \hat{\sigma}_z + \frac{\hbar g}{2}\left[A(t)\hat{\sigma}_++A^\star (t)\hat{\sigma}_-\right], \label{ham.gener}
\end{equation}
where $A(t)=\exp\left[\mathrm{i}\int_0^t f(\tau)\mathrm{d}{\tau}\right]$.

We perform a
Fourier-series expansion for the effective periodic transverse drive $A(t)$:
\begin{equation}
  A(t)=\sum_{m=-\infty}^\infty \Delta_m e^{i m \Omega t},
\end{equation}
where $\Delta_m$ are the sideband amplitudes.
For sinusoidal modulation with $f_{\rm sin} (t) = \delta \cos (\Omega t)$, the Jacobi-Anger relation~\cite{Stegun}
immediately gives $\Delta_{ m}^{\rm sin}=J_{m}(\delta/\Omega)$, where
$J_m(x)$ is the Bessel function of the first kind. For the
periodic latching modulation with $f_{sq}(t)=\delta\mathop{\rm sgn}\left[\cos(\Omega
t)\right]$, the driving amplitude has a piecewise representation
\begin{eqnarray}
A_{\rm sq}(t)=e^{i\int_0^t f_{\rm sq}(\tau)d\tau}=\left\{
\begin{array}{rl}
e^{i\delta \left(t- \frac{2\pi k}{\Omega} \right)} & {\rm if\ } -\frac{\pi}{2\Omega} +  \frac{2\pi k}{\Omega}< t
<\frac{\pi}{2\Omega}+ \frac{2\pi k}{\Omega},\\
 e^{-i\delta\left[t - \frac{\pi(2k+1)}{\Omega} \right]}  &  {\rm if\ } \frac{\pi}{2\Omega} +  \frac{2\pi k}{\Omega}< t <\frac{3\pi}{2\Omega}+  \frac{2\pi k}{\Omega},
\end{array} \right.
\end{eqnarray}
where $k$ is an integer, from which we obtain the sideband amplitude for periodic latching modulation,
\begin{eqnarray}
\Delta^{\rm sq}_m=\frac{\Omega}{2\pi}\int_{0}^{\frac{2\pi}{\Omega}}
e^{-im\Omega t}A_{\rm sq}(t)dt
=\frac{2}{\pi}\frac{\Omega\delta}{\Omega^2m^2-\delta^2}\sin\left(\frac{\pi m}{2}-\frac{\pi\delta}{2\Omega}\right). \label{exactRWA}
\end{eqnarray}
Note that the sideband amplitudes of both
sinusoidal modulation and latching modulation transform in the same way under $m\rightarrow -m$, namely $\Delta^{\rm sq}_{-m} = (-1)^{m} \Delta^{\rm sq}_{m}$ and
$\Delta^{\rm sin}_{-m} = (-1)^{m} \Delta^{\rm sin}_{m}$.


The resulting effective Hamiltonian written in a frame rotating at frequency $\omega_{0}-\omega$ around the $z$-axis becomes
\begin{equation}
  \hat{H}_{\rm eff}(t)=\frac{\hbar}{2}\left[\sum_{m=-\infty}^\infty g\Delta_m^{\rm sq} \mathrm{e}^{\mathrm{i} (m \Omega -\omega +\omega_{0}) t}  \hat{\sigma}_{+}+\textrm{H.c.}\right].\label{withapprox}
\end{equation}
When $\omega\approx \omega_0+m\Omega$, and if the other driving fields are not too strong $g|\Delta_n^{\rm sq}|<\Omega$, $n\neq m$, we can make the rotating-wave approximation by neglecting the non-resonant driving fields. When $\Omega$ is small, the RWA results can be improved by adding Bloch-Siegert and higher order corrections (so-called generalized Bloch-Siegert shift~\cite{Tuorila10}).

We find the steady state occupation probability $P_e$ by solving the Lindblad form master equation analytically around every resolvable resonance~\cite{EberlyAllen87, motional}. The master equation including the pure dephasing and the energy relaxation processes, with rates $\Gamma_\varphi$ and $\Gamma_1$, respectively, is written for two lowest transmon levels using the Hamiltonian (\ref{ham.first})
\begin{equation}
\frac{\mathrm{d}{\hat{\rho}(t)}}{\mathrm{d}{t}}=-\frac{\mathrm{i}}{\hbar}[\hat{H}_{\rm eff}(t),\hat{\rho}(t)]-\frac{1}{4}\Gamma_\varphi\left[\hat{\sigma}_z,\left[\hat{\sigma}_z,\hat{\rho}\right]\right]
 +\frac{1}{2}\Gamma_{1}\left(2\hat{\sigma}_-\hat{\rho}\hat{\sigma}_+-\hat{\sigma}_+
\hat{\sigma}_-\hat{\rho}-\hat{\rho}\hat{\sigma}_+\hat{\sigma}_-\right).
\end{equation}
Note that when decoherence is introduced, the widths of the sidebands are broadened due to both the decoherence rate $\Gamma_2=\Gamma_1 /2+\Gamma_\varphi$ and the power broadening caused by the strong transverse driving, $g\Delta^{\rm sq}_m$ ~\cite{EberlyAllen87}, yielding a linewidth $\lesssim\sqrt{\Gamma_2^2+(g\Delta_m^{\rm sq})^2\Gamma_2/\Gamma_1}$. By adding the contributions from all resolvable resonances, we get
\begin{equation}
  P^{\rm sq}_e=\sum_{m=-\infty}^\infty \frac{\frac{\Gamma_2}{2\Gamma_1} \left( g\Delta_m^{\rm sq}\right)^2}{\Gamma_2^2+(\omega_0-\omega+m\Omega)^2+\frac{\Gamma_2}{\Gamma_1}\left( g\Delta_m^{\rm sq}\right)^2} \label{Pe_sq}
\end{equation}
for the steady state occupation probability of the qubit excited state.

\begin{figure*}
\centering
\includegraphics[width=0.9\linewidth]{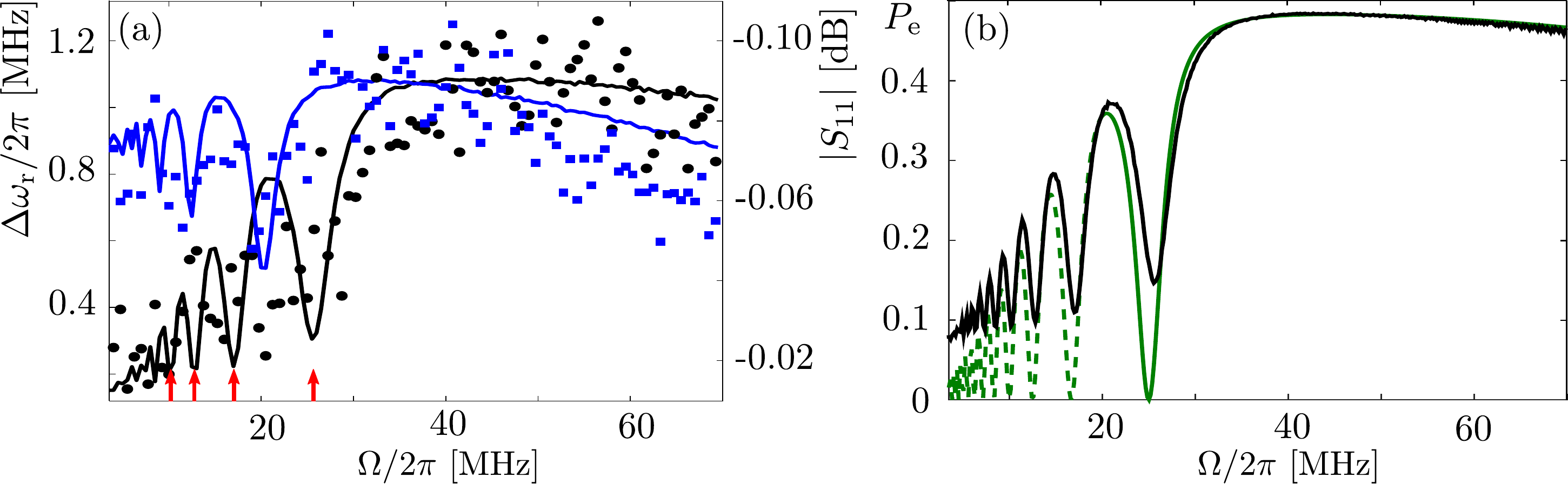}
\caption{\label{comparison}
Population and frequency shifts along the second sideband. (a) Comparison between sinusoidal (blue) and periodic latching (black) modulation, with the same measurement parameters and the same modulation amplitude as in Fig.~\ref{Square_fig}. The axis on the left represents the total dispersive shift $\Delta \omega_{r}$ of the cavity, as predicted by the numerical theoretical simulation (continuous lines), see Eq.~(\ref{shift}). The right axis shows the measured change in the microwave reflection coefficient $|S_{11}|$ (data points), referenced to the value corresponding to the ground state of the qubit. The vertical (red) arrows indicate the position of  antiresonances (coherent destruction of tunneling) predicted by Eq. (\ref{antiresonance}). (b) Comparison between the numerical simulation (continuous black line) and the RWA approximation for
periodic latching modulation, leading to analytical results Eq. (\ref{exactRWA}) and Eq. (\ref{Pe_sq}) (green line) for the population $P_{e}$ of the first excited state of the two-level transmon. The parameters are as in Fig.~\ref{Square_fig} except here $T=0$ mK.  At small values of $\Omega$ the RWA is represented as a dashed line to indicate that it is outside its expected range of validity.
}
\end{figure*}

In Fig. \ref{comparison} (b) we present the results of the RWA method for the second sideband $m=2$. In this figure we use  Eq. (\ref{Pe_sq}) with
the sideband amplitude for latching modulation given by Eq. (\ref{exactRWA}).
We find that the results of the RWA are in reasonably good agreement with the numerical ones down to $\Omega /2\pi \approx 20 $ MHz (continuous green line). Below this value (dashed green line) we see deviations as the rotating-wave approximation becomes inadequate for reproducing the numerical data. Nevertheless, the RWA predicts relatively well the position of the resonances.

The RWA analysis explains easily some of the features of the periodic latching modulation spectrum and the sinusoidal spectrum noted in the previous subsection. Firstly, note that in the limit $\Omega \gg \delta$ the two spectra will both eventually drop to zero if $m\neq 0$ or saturate to $\Delta_{0}^{\rm sq}|_{\Omega \gg \delta} = \Delta_{0}^{\rm sin}|_{\Omega \gg \delta} =1$ if $m=0$, which follows immediately from $\Delta_{ m}^{\rm sin}=J_{m}(\delta/\Omega)$ and the result for $\Delta_{ m}^{\rm sq}$ from Eq. (\ref{exactRWA}). This can be understood as a consequence of the time-energy uncertainty principle and of motional averaging~\cite{motional}: the energy levels corresponding to the latching points cannot be discriminated anymore if the time $\pi/\Omega$ that the system spends at each of these points is much shorter that the energy difference $2\hbar\delta$. Secondly, for $\delta/\Omega=|m|\neq 0$ we get from Eq. (41) that $|\Delta_m^{\rm sq}|=1/2$. This means that for the latching modulation all the sidebands have the same amplitude at $\omega-\omega_0= m \Omega=\pm \delta$ for all $m \neq 0$. This property does not hold for the sinusoidal modulation, since
$J_{m}(|m|)$
take different values depending on $m$. Finally, using the RWA we can understand why the first maximum of the periodic latching modulation is shifted towards higher values compared to the sinusoidal, as noted in the previous subsection, see Fig.~\ref{comparison} (a). Indeed, the RWA is accurate around the region of the first maxima, therefore one can simply analyze the maxima of the exact results for sinusoidal and the latching modulation. Qualitatively, the existence of a shift originates in the rescaling of the argument $\delta /\Omega$ of the Bessel-function solution for sinusoidal modulation by $\pi/2$ in Eq. (\ref{exactRWA}).

\section{Conclusions}
\label{sec:conclusions}
We have shown that the periodic latching modulation of the transition frequency of a qubit is conceptually different from the continuous drive forms used in the conventional studies of LZS-interference. We have adapted the adiabatic-impulse method for the case of abrupt and periodic switching, and the results are shown to be in good agreement with more elaborate numerical calculations. We have studied the periodic latching modulation experimentally by employing a transmon with flux bias modulated with a square pulse pattern. We measured a spectrum where two regimes (slow-modulation and fast-modulation) can be clearly distinguished. The spectrum has a rich structure of sidebands, due to resonances and anti-resonances (coherent destruction of tunneling).
The experimental data were in good agreement with our theoretical models, and we were able to extract the information about the pulse shape from the region of low modulation frequency.
Our results open the way for simulating various forms of dephasing noise and for realizing experiments where the switching of the qubit frequency has a specific, non-sinusoidal time dependence.

\acknowledgments

We are very grateful to Mika Sillanp\"a\"a and to Pertti Hakonen for many useful discussions at various stages of this project and to Juha Pirkkalainen for assistance with sample design and fabrication. We also thank W. C. Chien for help with the experiments. This work was supported in part by the facilities and staff of the Yale University Faculty of Arts and Sciences High Performance Computing Center and used the cryogenic facilities of the Low Temperature Laboratory at Aalto University. We acknowledge financial support from Army Research Office W911NF1410011, NSF DMR-1301798, the Magnus Ehrnrooth Foundation, the Finnish Academy of Science and Letters, FQXi, the Academy of Finland (projects 263457, 135135), and the Center of Excellence ``Low Temperature Quantum Phenomena and Devices'' (project 250280).

\appendix*

\section{Validity of the sudden approximation}
\label{AppendixA}

Consider a time-dependent Hamiltonian $\hat{H}(t)$ that changes rapidly during the time-interval $[0,T]$. The time evolution during this interval can be written as an iterative solution:
\begin{equation}
  \hat{U}_T=1-\frac{\mathrm{i}}{\hbar} \int_0^T \hat{H}(\tau)d{\tau} -\frac{1}{\hbar^2} \int_0^T d{\tau} \int_0^\tau d{\tau'} \hat{H}(\tau) \hat{H}(\tau')+... \label{iter.U}
\end{equation}
Suppose now that the system starts in the initial state $|\psi_{\rm r}^-\rangle$, the lower eigenstate on the right side of the avoided crossing. Then, in the spirit of Ref.~\onlinecite{Messiah}, we introduce the measure for the validity of the sudden approximation:
\begin{equation}
  w^{(r)}=1 - \left|\langle \psi^{(r)}_{-}|\hat{U}_T |\psi^{(r)}_{-}\rangle\right|^2 =\left|\langle \psi^{(r)}_{+}|\hat{U}_T |\psi^{(r)}_{-}\rangle\right|^2. \label{w.def}
\end{equation}
This is the probability that the time-evolution during the sudden change of Hamiltonian brings the initial state $|\psi^{(r)}_-\rangle$ into the orthogonal state $|\psi^{(r)}_+\rangle$, with $\hat{I} = |\psi^{(r)}_-\rangle \langle \psi^{(r)}_- | + |\psi^{(r)}_+\rangle \langle \psi^{(r)}_+ |$. For an  instantaneous $\hat{U}_{\rm T}$ the state remains unchanged $\hat{U}_{\rm T}|\psi^{(r)}_-
\rangle = |\psi^{(r)}_- \rangle$ and $w^{\rm (r)}=0$.

By using the expansion (\ref{iter.U}), we obtain for  Eq. (\ref{w.def}) in the second order in $\hat{H}$
\begin{equation}
  w^{(\rm r)}=\frac{1}{\hbar^2}\left< \psi^{(r)}_- \left|\int_0^T \hat{H}(\tau)d{\tau}  \int_0^T \hat{H}(\tau)d{\tau}\right| \psi^{(r)}_-\right> -\frac{1}{\hbar^2}\left< \psi^{(r)}_-\left| \int_0^T \hat{H}(\tau)d{\tau}\right| \psi^{(r)}_-\right>^2, \label{W.2}
\end{equation}
and further,
\begin{equation}
w^{(r)}=\frac{T^2}{\hbar^2}\left[\left< \psi^{(r)}_- \left|
\bar{H}^2 \right| \psi^{(r)}_-\right>
-\left< \psi^{(r)}_- \left|\bar{H}\right| \psi^{(r)}_-\right>^2 \right] =\frac{T^2}{\hbar^2} (\Delta \bar{H})^2 ,
\end{equation}
where $\bar{H}=\frac{1}{T} \int_0^T \hat{H}(\tau)d{\tau}$ is the time-averaged Hamiltonian during the time evolution $T$. Note that the transition probability depends only on the averaged Hamiltonian $\bar{H}$.

For our system, the Hamiltonian is
\begin{equation}
\hat{H}(t)=\frac\hbar 2[\nu + f_{\rm sq}(t)]\hat{\sigma}_z  + \frac{\hbar g}{2}\hat{\sigma}_x.
\end{equation}
where $f_{\rm sq}(t)$ ideally describes the periodic latching modulation with an instantaneous switching events, see the diagram in Fig. \ref{scheme} (a). However, in practice the ramp between the two latches has a finite raise/fall time $T$.
Let us use the parametrization for the Hamiltonian at $t=0$ corresponding to the right side of the crossing
\begin{equation}
\hat{H}^{(r)}(0) = |\epsilon_{\pm}^{(r)}| \left(\begin{array}{cc} \cos\theta^{(r)} & \sin\theta^{(r)} \\ \sin\theta^{(r)} & -\cos\theta^{(r)}
\end{array} \right),
\end{equation}
where $\epsilon_{\pm}^{(r)} = \pm \frac{\hbar}{2}\sqrt{(\nu + \delta)^2 + g^2}$ and
$\nu + \delta = \sqrt{(\nu + \delta)^2 + g^2}\cos\theta^{(r)}$,
$g = \sqrt{(\nu + \delta)^2 + g^2}\sin\theta^{(r)}$.
The eigenstates $|\psi^{(r)}_\pm\rangle$ are
\begin{equation}
|\psi^{(r)}_-\rangle = \left(\begin{array}{c} -\sin\frac{\theta^{(r)}}{2} \\ \cos\frac{\theta^{(r)}}{2}
\end{array}\right), ~~
|\psi^{(r)}_+\rangle = \left(\begin{array}{c} \cos\frac{\theta^{(r)}}{2} \\ \sin\frac{\theta^{(r)}}{2}
\end{array}\right),
\end{equation}
with $\hat{H}^{(r)}(0)|\psi^{(r)}_\pm\rangle = \epsilon_{\pm}^{(r)}|\psi^{(r)}_\pm\rangle$.
The angle $\theta^{(r)}$ is found from
\begin{eqnarray}
\cos\theta^{(r)} &=& \frac{\nu + \delta}{\sqrt{(\nu + \delta )^2+ g^2}}, \\
\sin\theta^{(r)} &=& \frac{g}{\sqrt{(\nu + \delta )^2+ g^2}}.
\end{eqnarray}
Similarly, on the left side the Hamiltonian can be diagonalized and the eigenvalues and eigenfunctions are obtained from the right-side values by replacing $\delta$ with $-\delta$,
\begin{eqnarray}
\epsilon_{\pm}^{(\ell)}&=&\left.\epsilon_{\pm}^{(r)}\right|_{\delta\rightarrow -\delta},\\
|\psi^{(\ell)}_\pm\rangle &=& \left.|\psi^{\rm (\ell)}_\pm\rangle \right|_{\delta\rightarrow -\delta}.
\end{eqnarray}
Then the elements of the matrix  Eq. (\ref{matrixU}) are $\sqrt{p_{\rm s}}\equiv \sin [(\theta^{(\ell)}-\theta^{(r)})/2]$ and
$\sqrt{1-p_{\rm s}}\equiv \cos [(\theta^{(\ell)}-\theta^{(r)})/2]$. Explicitly,
$|\psi^{(\ell)}_+\rangle = \sqrt{1-p_{\rm s}}|\psi^{(r)}_+\rangle +
\sqrt{p_{\rm s}}|\psi^{(r)}_-\rangle$ and
$|\psi^{(\ell)}_-\rangle = -\sqrt{p_{\rm s}}|\psi^{(r)}_+\rangle +
\sqrt{1-p_{\rm s}}|\psi^{(r)}_-\rangle$. Note that we have $g>0$ and $\theta^{(r,\ell)}\in [0,\pi]$, therefore
$\sqrt{p_{\rm s}}$ and $\sqrt{1-p_{\rm s}}$ come out indeed positive, as used in the main text.

During the ramp time $T$
the change in the detuning
$f(t)=(2t/T -1 )\delta $ can be assumed linear from $-\delta$ to $\delta$:
The time-averaged Hamiltonian is
\begin{eqnarray}
\bar{H}^2&=&\frac{\hbar^2}{4}\left[ \nu ^2+g^2\right] ,\\
  \bar{H}&=&\frac{1}{T} \int_0^T \hat{H}(\tau)d{\tau}=\frac\hbar 2 \left [ \nu \hat{\sigma}_z +g \hat{\sigma}_x\right].
\end{eqnarray}
\begin{figure}
\includegraphics[width=0.5\linewidth]{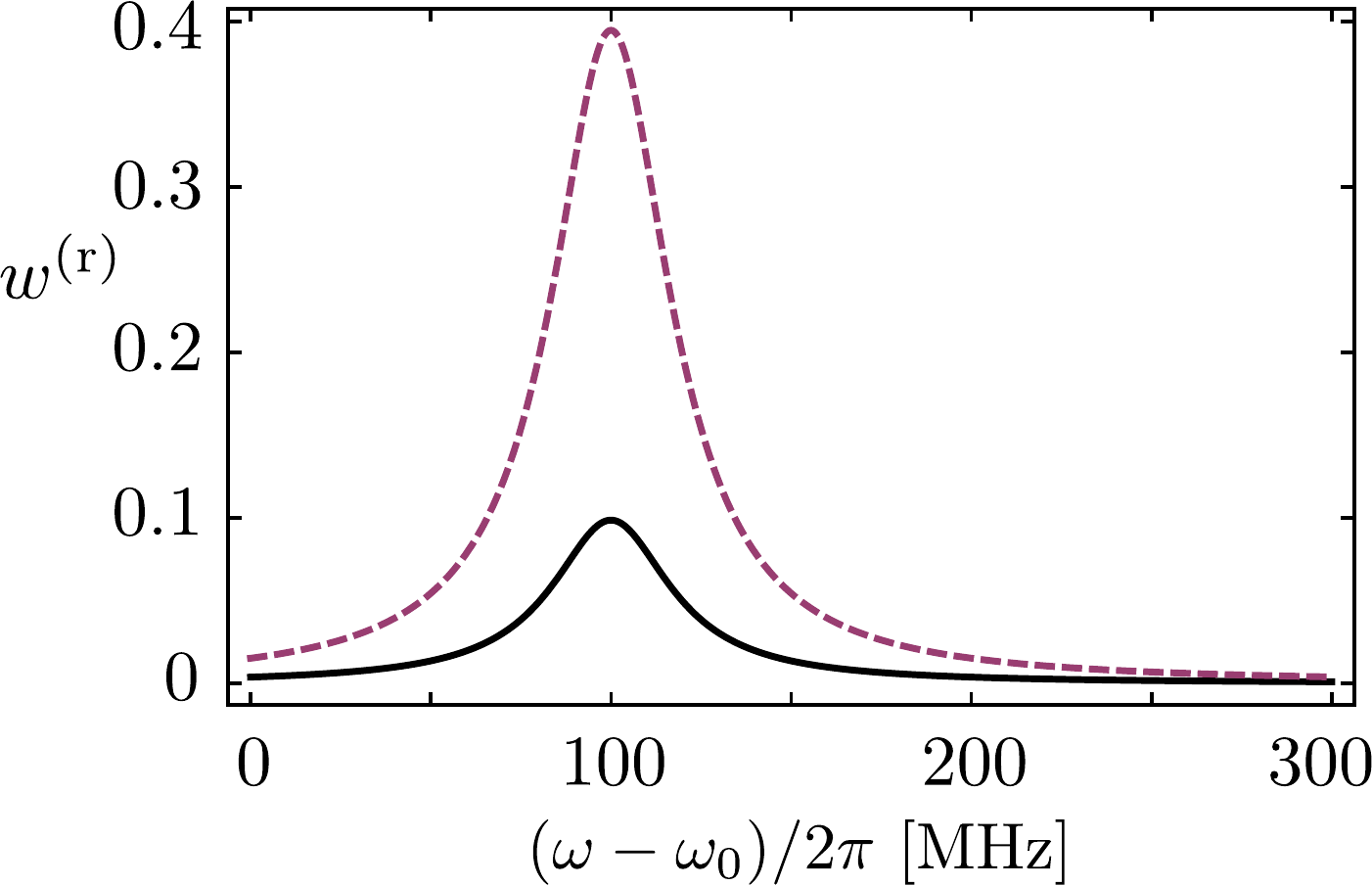}
\caption{Transition probability (\ref{W}) as a function of $\omega -\omega_{0}$, with values $g/2\pi=20$ MHz and $\delta/2\pi=100$ MHz for $T=2$~ns (dashed line) and $1$~ns (solid line). \label{fig.W}}
\end{figure}
With these specifications we get
\begin{eqnarray}
\left< \psi^{(r)}_- \left|
\bar{H}^2 \right| \psi^{(r)}_-\right> &=&\frac{\hbar^2}{4}\left[ \nu^2+g^2\right],\\
\left< \psi^{(r)}_- \left|\bar{H}\right| \psi^{(r)}_-\right>^2 &=&
-\frac{\hbar}{2}[\nu\cos\theta^{(r)} + g \sin\theta^{(r)}],
\end{eqnarray}
and
the transition probability (\ref{W.2}) during the linear ramp is
\begin{equation}
w^{(r)}=\frac{T^2}{4}[\nu \sin\theta^{(r)} -  g\cos\theta^{(r)} ]^2 =\frac{T^2\delta^2}{4}
\frac{g^2 }{g^2+(\nu +\delta)^2},\label{W}
\end{equation}
which is a Lorentzian peak around $\nu = \omega_0-\omega = - \delta$ with width $g$ and maximum value
\begin{equation}
w_{\rm max}=\frac{T^2 \delta^2}{4},\label{maxi}
\end{equation}
see Fig.~\ref{fig.W}. The same expression is obtained if one starts from the state $|\psi_{\rm r}^+\rangle$. If we consider the same sudden jump limit on the left side of the crossing, the result is
\begin{equation}
w^{(\ell)}=\frac{T^2 \delta^2}{4}\frac{g^2 }{g^2+(\nu -\delta)^2},\label{Wleft}
\end{equation}
yielding the same maximum value as in Eq. (\ref{maxi}). Thus, the sudden approximation is valid if the ramp time $T\ll \delta^{-1}$.
Experimentally, we estimate  $T=1-2$ ns, which for $\delta/2\pi = 100$ MHz
corresponds to $w_{\rm max}(T=1\text{ ns})=0.1$ and $w_{\rm max}(T=2\text{ ns})=0.4$ respectively, see Fig.~\ref{fig.W}. We conclude that the sudden approximation is fair when $\omega_0-\omega=\delta$, and improves fast when we move away from the resonance.

\end{document}